\documentclass[12pt]{article}
\usepackage{amsmath}
\usepackage{amsfonts}
\usepackage{graphics}
\usepackage{color}
\newcommand\NewTextA[1]{{\color[rgb]{0.35,0.,0.}#1}}
    \renewcommand\NewTextA[1]{{#1}}

\newcommand\Intd{{\mathrm{d}}}
\newcommand\Remove[1]{{\raise 0.9ex\hbox{$\!\times$}\hspace{-0.6em} \lower 0.15ex\hbox{$#1$}}}

\newcommand\VEV[1]{{\left<0\right|#1\left|0\right>}}
\newcommand\eqdef{{\,\stackrel{\scriptstyle\mathrm{def}}{=}\,}}
\newcommand\rme{{\mathrm{e}}}
\newcommand\rmi{{\mathrm{i}}}

\newcommand\Fdelta{{\underline{\delta}}}
\newcommand\Fint[1]{{\underline{\Intd^4#1}}}
\newcommand\Mass{{\mathsf{m}}}
\newcommand\MMass{{\mathsf{M}}}
\begin{document}

\centerline{\huge Multi--particle quantum fields for}\vspace{1ex}
\centerline{\huge bound states and interactions}

\vspace{6ex}\noindent\centerline{Peter Morgan}

\vspace{0.4ex}\noindent\centerline{Physics Department, Yale University}

\vspace{0.4ex}\noindent\centerline{peter.w.morgan@yale.edu}

\vspace{0.4ex}\noindent\centerline{\today}

\newcommand\Chi{{\mathsf{X}}}

\vspace{4ex}
\noindent\textbf{Abstract:} The Fock--Hilbert space generated by a single--particle interaction--free Wightman field is augmented by introducing non--trivial multi--particle (that is, multi--point, multilinear) quantum fields, which is justified insofar as Haag's theorem establishes that free field Fock--Hilbert spaces cannot model bound or interacting states.
Two Gaussian constructions are given: one that modifies the combinatoric factors and masses associated with products of propagators and a second for which locality is determined by the center of mass of the $n$--particles and relative separations of the $n$--particles determine the strength of resonance; it is shown how the two constructions may also be used together. 
Finally, a method is given for generating non-Gaussian $n$--particle quantum fields that quite closely tracks familiar interacting quantum fields but that is significantly better--defined and that offers a much richer algebraic structure for future use.

\vspace{4ex}\noindent\centerline{Keywords: Quantum Fields, Quantum Optics,
Bound and Interacting Particles.}

\newpage

\section{Introduction}\label{Introduction}
A possible reaction to Haag's theorem\cite{Fraser,EarmanFraser} is to think that the Fock--Hilbert space generated by free 1--particle Wightman fields\cite[Ch. II]{Haag} is not ``big enough'' to model bound states or interacting states generally, which is taken here to suggest adding multi--particle bound and interacting states explicitly as multi--point fields $\hat\phi(x_1, ..., x_n)$ and constructing physical states as superpositions of different products of 1--particle and multi--particle fields acting on the vacuum state.
We are already accustomed to introducing \NewTextA{quasiparticles and collective excitations} that are effectively bound multi--particle fields such as phonons in statistical physics and such as for protons in high energy physics.

We will first introduce here Gaussian scalar multi--particle fields, for which the two--operator Vacuum Expectation Value (VEV) $\VEV{\hat\phi(x_1,...,x_n)\hat\phi(x_1',...,x_n')}$ determines all VEVs.
The first construction, in \S\ref{AlmostTrivialNParticle}, is almost trivial, in that it constructs the two--operator $n$--particle VEV in terms of the two--operator 1--particle VEV $\VEV{\hat\phi(x)\hat\phi(x')}$,
$$\VEV{\hat\phi(x_1,...,x_n)\hat\phi(x_1',...,x_n')}=\prod_{j=1}^n\VEV{\hat\phi(x_j)\hat\phi(x_j')}.
$$
for which the $n$--particle operators in general commute whenever \emph{all} separations $x_i-x_i'$ are space--like.
States and measurements that use these operators allow combinatoric factors associated with the field to be modified.
For this and any construction of an $n$--particle field, we can symmetrize $\hat\phi(x_1, ..., x_n)$ as $\hat\phi^s(x_1, ..., x_n)=\sum_{\sigma\in S_n}\hat\phi(x_{\sigma(1)}, ..., x_{\sigma(n)})$, so that $\hat\phi^s(x_1, ..., x_n)$ is independent of the order of $x_1$,...$x_n$, and symmetrization will be a natural result of the construction of non--Gaussian $n$--particle fields in \S\ref{Polarization}.

The second construction, in its simplest form in \S\ref{TwoParticle} and \S\ref{nParticle}, introduces the center of mass of the $n$ particles, $X$, and the $n-1$ relative separations $z_j=x_{j+1}-x_j$, in such a way that the locality of $\hat\phi(x_1,...,x_n)$ is determined exclusively by the center of mass.
In this case, $\hat\phi_{z_1,...,z_{n-1}}(X)=\hat\phi(x_1,...,x_n)$ is essentially a $1$--particle Wightman field localized at $X$, for which $z_1$,...,$z_{n-1}$ are continuous indices, however the construction here is simple when Lorentz invariantly expressed in terms of the relative separations $z_j$ instead of being simple when expressed in terms of finite--dimensional or continuous irreducible representations of the Lorentz group.

When introducing structure additional to that of Wightman fields, we will aim to satisfy the spirit of the Haag--Kastler axioms\cite[\S III.1]{Haag}, that there is a Poincar\'e covariant Hilbert space and a Poincar\'e invariant vacuum state, satisfying the Spectrum Condition, Additivity, Hermiticity, Locality, and Completeness, which do not \NewTextA{explicitly} require that fields must be 1--point, 1--particle objects (we will henceforth refer to $n$--particle fields, not to $n$--point fields or $n$--linear fields, notwithstanding any preference for the latter, or at least for the latter two).

We will find that Additivity is \emph{not} generally satisfiable by a multi--particle construction, however Additivity is more a theoretical preference for the sake of simplicity than an empirical principle and is not infrequently omitted from relatively informal discussions of axiomatic approaches to quantum field theory (as in \cite{FredenhagenEtAl}, for example); multi--particle fields, as operator--valued distributions, instead satisfy more--or--less natural extensions of the linearity of Wightman fields.

The construction in \S\ref{RealScalarFields} of $n$--particle fields for which commutativity is determined by only whether the centers of mass of two sets of $n$ points are space--like separated, is extended in \S\ref{MultipleCenterOfMass} to $n$--particle fields for which commutativity is determined by several differently weighted centers of mass $X_1$, ..., $X_m$, distinguishing several components of the whole, each of which is contained in the convex hull of the $n$ points.
This construction suggests a fairly reasonable way to restrict Locality: the Haag-Kastler axioms associates an algebra of operators with any region of space--time, and requires algebras associated with space-like separated regions to commute, however we will adopt a more--or--less minimal and natural restriction for an $n$--particle operator, that space--time regions are required to be convex.
The success of this construction corresponds to requiring that algebras of operators are associated with convex regions of space--time because the multiple centers of mass of $n$ points are located within the convex hull of those points.
The importance given here to the center of mass conforms well both with its importance in classical physics and with its importance when modeling quasiparticles and collective excitations as single quantum particles such as phonons; where we construct an $n$--particle field as composed of several interacting components, we expect the centers of mass of the several components to be significant.

As a larger Hilbert space than the Fock--Hilbert space of the Wightman field, a significantly larger range of models is available, which can only be more empirically capable than a free quantum field; given that quantum optics usefully applies just the quantized free electromagnetic field, we can credibly hope that including multi--particle fields might allow more widely useful application, whether or not such constructions might be thought fundamental.

$n$--particle fields require a rather different way of thinking about quantum states.
We have become accustomed to discussing the evolution of states of a free quantum field, so that a 2--particle input state $\tilde\phi_{\mathrm{in}}(k_1)\tilde\phi_{\mathrm{in}}(k_2)|0\rangle$, with momenta $k_1$, $k_2$, evolves over an asymptotically long period to include, say, a 3--particle output component (of a different type of particle)
$\tilde\xi_{\mathrm{out}}(k'_1)\tilde\xi_{\mathrm{out}}(k'_2)\tilde\xi_{\mathrm{out}}(k'_3)|0\rangle$.
Given 2-- and 3--particle field operators, however, we may take a prepared state to have always included the possibility of 2-- and 3--particle components, as a superposition of vector states,
$$\left[\alpha_1\tilde\phi(k_1)\tilde\phi(k_2)+\alpha_2\tilde\phi(k_1,k_2)+\beta_1\tilde\xi(k'_1)\tilde\xi(k'_2)\tilde\xi(k'_3)
+\beta_3\tilde\xi(k'_1,k'_2,k'_3)\right]|0\rangle,$$
say (with no distinction between input and output Hilbert spaces now required), or perhaps as a mixture of such vector states.
The probabilities of detecting the 1--, 2-- and 3--particle components vary differently with changes of separation, with an effective dynamics determined by the construction of the $n$--particle fields.

The construction of multi--particle real scalar fields in \S\ref{RealScalarFields} is extended to Dirac and Maxwell fields in \S\ref{DiracMaxwellFields}.
Finally, \S\ref{Polarization} introduces a means for generating non--Gaussian $n$--particle quantum fields that is capable of quite closely approaching familiar interacting quantum fields when just Gaussian 1--particle quantum fields are used.
Significantly, however, the generation of non--Gaussian $n$--particle quantum fields allows a much richer and better--defined construction when we use the Gaussian $n$--particle quantum fields that we have introduced in \S\ref{RealScalarFields} and \S\ref{DiracMaxwellFields}.
The introduction of this conceptually different form offers a step away from the difficulties of regularization and renormalization that Lagrangian and Hamiltonian deformation entail.

Wightman field theory, and constructive quantum field theory more generally, has apparently not previously been extended towards multi--particle fields (see \cite[\S 3.4]{Streater}, for example, for a 1975 critique of the Wightman axioms; see \cite{Summers}, for example, for a recent overview of constructive quantum field theory).
There have instead been emphases, for example, on 1+1-- and 1+2--dimensional models, Euclidean models, wedge--localization, and deformation approaches.
Interacting quantum fields outside of axiomatic traditions have been overwhelmingly concerned with Lagrangian dynamics, regularization, and renormalization.

\section{Multi--particle real scalar fields}
\label{RealScalarFields}
After a very brief discussion of a non--interacting 1--particle real scalar field, we introduce an almost trivial multi--particle field in \S\ref{AlmostTrivialNParticle}.
The specific case of $n$--particle center of mass localized fields introduces new possibilities, discussed for two particles in \S\ref{TwoParticle}, for $n$ particles in \S\ref{nParticle}, and generalized to an $n$--particle multiple center of mass field in \S\ref{MultipleCenterOfMass}.
Finally for the real scalar field case, invariance under symmetries is discussed in \S\ref{InvarianceUnderSymmetry}.

\subsection{The familiar 1--particle field}
A non--interacting real scalar field on a 4--dimensional Minkowski space can be presented in a manifestly Lorentz covariant way as
\begin{eqnarray*}
  &&\hat\phi(x)=\int \left[a(k)\rme^{-\rmi k\cdot x}+a^\dagger(k)\rme^{\rmi k\cdot x}\right]
                           \Fint{k},\qquad a(k)|0\rangle=0, \hspace{7em}\cr
  &&\VEV{\hat\phi(k)\hat\phi(k')}=[a(k),a^\dagger(k')]=\Fdelta^4(k-k')\rmi\tilde\Delta_{+\Mass}(k)\cr
  && \rule{0pt}{2.75ex}\hspace{13.5em}  =\Fdelta^4(k-k')\Fdelta(k\!\cdot\!k-\Mass^2)\theta(k_0)
\end{eqnarray*}
\mbox{\Large [}notation, to avoid proliferating factors of $2\pi$: $\Fdelta(k)=2\pi\delta(k)$ and
$\Fint{k}\!=\,$\raisebox{2pt}{\scriptsize $\displaystyle\frac{\textstyle\Intd^4k}{\textstyle(2\pi)^4}$}\mbox{\Large ]}.
The real--space commutator
$$
  [\hat\phi(x),\hat\phi(x')]=\int 2\rmi\sin{\!\Bigl(\!k\!\cdot\!(x'-x)\!\Bigr)}\rmi\tilde\Delta_{+\Mass}(k)\Fint{k}
$$
is zero if $x'-x$ is space--like, satisfying Locality, because the integrand is odd under the space--like reflection $k\rightarrow k-2\frac{k\cdot(x'-x)}{(x'-x)\cdot(x'-x)}(x'-x)$.
Critically for a quantum field, the commutator $[a(k),a^\dagger(k')]$ is a positive semi--definite infinite--dimensional matrix that is diagonal in the wave--number indices $k,k'$, so that we can use the vacuum state to construct a Fock--Hilbert space, and the factor $\rmi\tilde\Delta_{+\Mass}(k)$ projects to positive energy in all inertial frames.
Although the 4--momentum operator $P_\mu$ is often constructed as a function of quantum fields, we will here directly construct it as a generator of active translations,
$$P_\mu\hat\phi(x_1)\cdots\hat\phi(x_n)|0\rangle=\left.\frac{1}{\rmi}\frac{\partial}{\partial z^\mu}
   \hat\phi(x_1+z)\cdots\hat\phi(x_n+z)|0\rangle\right|_{z=0},\qquad P_\mu|0\rangle=0,
$$
which admits straightforward extension to the multi--particle case.

It is worthwhile to keep in mind that quantum fields are operator--valued distributions, so that only quantum fields ``smeared'' by test functions taken from some test function space are operators that act on the Hilbert space, where the test function space is usually taken to be a Schwartz space of functions that are smooth both in real space and in momentum space.
The creation and field operators of a quantum field theory can be written as
$$
  a_f^\dagger=\!\int\!a^\dagger(k)\tilde f(k)\Fint{k},\qquad
  \hat\phi_f=a_{f^*}+a_f^\dagger=\!\int\!\hat\phi(k)\tilde f(k)\Fint{k},
$$
so that $\VEV{\hat\phi_f^\dagger\hat\phi_g}=[a_f,a^\dagger_g]=(f,g)\eqdef\int \tilde f^*(k)\rmi\tilde\Delta_{+\Mass}(k)\tilde g(k)\Fint{k}$ is a positive semi--definite inner product on the test function space that projects to positive frequency components of the test functions, and $[\hat\phi_f,\hat\phi_g]=(f^*,g)-(g^*,f)$ is zero whenever the test functions $f$ and $g$ have space--like separated supports.
$\hat\phi_f^\dagger=\hat\phi_{f^*}$ is self-adjoint only if $f$ is real--valued, $f=f^*$.

\subsection{An almost trivial multi--particle field}
\label{AlmostTrivialNParticle}
The simplest example of a multi--particle field is the real scalar $n$--particle field
\begin{eqnarray*}
  &&\hat\phi(x_1, ..., x_n)=\int \left[a(k_1, ..., k_n)\rme^{-\rmi \sum k_i\cdot x_i}
                         +a^\dagger(k_1, ..., k_n)\rme^{\rmi \sum k_i\cdot x_i}\right]
                            \prod\limits_i\Fint{k_i},\cr
  &&\VEV{\hat\phi(k_1, ..., k_m)\hat\phi(k'_1, ..., k'_n)}
        =[a(k_1, ..., k_m),a^\dagger(k'_1, ..., k'_n)]\cr
  && \hspace{6em}=\delta_{m,n}\prod\limits_i\Fdelta^4(k_i-k'_i)\rmi\tilde\Delta_{+\Mass}(k_i),
                 \qquad a(k_1, ..., k_n)|0\rangle=0,\cr
  &&\VEV{\hat\phi(x_1, ..., x_m)\hat\phi(x'_1, ..., x'_n)}
       =\delta_{m,n}\prod\limits_i\rmi\Delta_{+\Mass}(x_i-x'_i).
\end{eqnarray*}
The multi--particle field $\hat\phi(x_1, ..., x_n)$ satisfies a natural extension of Locality, that the commutator $[\hat\phi(x_1, ..., x_n),\hat\phi(x'_1, ..., x'_n)]$ is zero if every $x_i-x'_i$ is space--like, which is consistent with Haag--Kastler Locality if we take the algebra of operators associated with a region $\mathcal{O}$ to be presented in its bounded Weyl form and smeared by smooth test functions $f(x_1, x_2, ..., x_n)$ that have support in $\mathcal{O}$ for each $x_i$.
We work with unbounded operators $\hat\phi_f$ instead of with bounded operators $\rme^{\rmi\hat\phi_f}$, but otherwise, in the spirit of the other Haag--Kastler axioms, Hermiticity, the Spectrum Condition, Poincar\'e covariance, and the construction of a Hilbert space are satisfied by construction as they are for Wightman fields by the action of creation operators on a vacuum state that is a zero eigenstate of all annihilation operators.
\NewTextA{Additivity ---that the algebra associated with $\mathcal{O}_1\cup\mathcal{O}_2$ is generated by the algebras associated with $\mathcal{O}_1$ and $\mathcal{O}_2$, $\mathcal{A}(\mathcal{O}_1\cup\mathcal{O}_2)=\mathcal{A}(\mathcal{O}_1)\vee\mathcal{A}(\mathcal{O}_2)$--- is not satisfiable by any multi--point formalism, because a function $f$ with $\mathrm{Supp}(f)\subseteq\left((\mathcal{O}_1\cup\mathcal{O}_2)\times(\mathcal{O}_1\cup\mathcal{O}_2)\right)$ cannot in general be generated as a sum of functions $f_i$ with, for each $i$, either $\mathrm{Supp}(f_i)\subseteq\mathcal{O}_1\times\mathcal{O}_1$ or $\mathrm{Supp}(f_i)\subseteq\mathcal{O}_2\times\mathcal{O}_2$, however an $n$--point linearity, with an algebra generated by $n$--point distributions and test functions with support in $\mathcal{O}^{\times n}$, is a natural replacement for Additivity.}
In terms of the smeared $n$--particle creation operators
$$a^\dagger_f=\sum\limits_n\int a^\dagger(k_1, ..., k_n)
                                               \tilde f_n(k_1, ..., k_n)\Fint{{}^n\hspace{-0.1em}k},$$
the corresponding annihilation operators, and the operator $\hat\phi_f=a_{f^*}+a_f^\dagger$, the commutation relations can be presented as for the 1--particle case as $[a_f,a^\dagger_g]=(f,g)$, where $(f,g)$ is required to be a positive semi--definite inner product on a much extended test function space, and as $[\hat\phi_f,\hat\phi_g]=(f^*,g)-(g^*,f)$.

The introduction of this kind of multi--particle field is non--trivial, but in the first instance the only change that it causes is a modification of combinatoric factors, because all propagators are of the form $\mathrm{i}\tilde\Delta_{+\Mass}(k)$.
If we introduce a 2--particle field smeared by a test function $f(x_1,x_2)=f_1(x_1)f_2(x_2)$, for example, denoted by  $\hat\phi_{f_1\otimes f_2}$, for which the $2\!\times\!2$--particle Vacuum Expectation Value (VEV) is $\VEV{\hat\phi_{f_3\otimes f_4}\hat\phi_{f_1\otimes f_2}}=(f_3^*,f_1)(f_4^*,f_2)$, as well as the usual $\hat\phi_f$, smeared by one 1--particle test function, we can construct a transition amplitude such as
\begin{eqnarray*}
  &&\VEV{[\hat\phi_{f_3}\hat\phi_{f_4}+\alpha\hat\phi_{f_3\otimes f_4}]
        [\hat\phi_{f_1}\hat\phi_{f_2}+\alpha\hat\phi_{f_1\otimes f_2}]}\cr
  &&\hspace{3em}=(1+\alpha^2)(f_3^*,f_1)(f_4^*,f_2)+(f_3^*,f_2)(f_4^*,f_1)+(f_3^*,f_4)(f_1^*,f_2),\hspace{3em}
\end{eqnarray*}
using superpositions of different types of 2--particle components.

A significant difference for multi--particle fields, however, is that a variety of masses may occur that may not occur for lower--degree multi--particle fields (including 1--particle fields), which effectively models the energy dependency of the multi--particles on space--time separations.
In non--scalar field cases, the masses allowed for multi--particle components may also depend on relative configurations of internal degrees of freedom.

\subsection{The $2$--particle center of mass localized field}
\label{TwoParticle}
An annihilation--creation commutator for the 2--particle case,
$[a(k_1, k_2),a^\dagger(k'_1, k'_2)]$, must be a positive semi--definite infinite--dimensional matrix, which to ensure translation invariance must be diagonal in the wave--number indices $k_1+k_2,k'_1+k'_2$.
We consider an example of the form
$$
  [a(k_1, k_2),a^\dagger(k'_1, k'_2)]=\Fdelta^4(k_1+k_2-k'_1-k'_2)
                   \rmi\tilde\Delta_{+\MMass}(k_1+k_2)\tilde M(k_1-k_2;k'_1-k'_2),
$$
which is invariant under a space--like reflection
$k_1+k_2\rightarrow k_1+k_2-2\frac{(k_1+k_2){\cdot}V}{V{\cdot}V}V$, for $V$ space--like and,
separately, we require invariance of $\tilde M(k_1-k_2;k'_1-k'_2)$ under the simultaneous reversals
$k_1-k_2\rightarrow -(k_1-k_2)$, $k'_1-k'_2\rightarrow -(k'_1-k'_2)$.
$k_1+k_2=k'_1+k'_2$ must be forward--pointing time--like to satisfy the Spectrum Condition,
but $k_1$, $k_2$, $k'_1$, and $k'_2$ do not individually have to be.
The infinite--dimensional matrix $\tilde M(k;k')$ is required to be positive semi--definite.

This 2--particle construction satisfies Locality insofar as the integrand in the field commutator
\begin{eqnarray*}
  [\hat\phi(x_1,x_2),\hat\phi(x'_1,x'_2)]=\int 2\rmi\sin{(k'_1\!\cdot\! x'_1+k'_2\cdot\! x'_2
                                                                      -k_1\!\cdot\! x_1-k_2\!\cdot\! x_2)}\hspace{3em}\cr
        \times\ [a(k_1, k_2),a^\dagger(k'_1, k'_2)]\Fint{k^{\,}_1}\,\Fint{k^{\,}_2}\,\Fint{k'_1}\,\Fint{k'_2}
\end{eqnarray*}
includes the factor $\sin{(k'_1\!\cdot\! x'_1+k'_2\cdot\! x'_2-k_1\!\cdot\! x_1-k_2\!\cdot\! x_2)}$, for which, because the momentum space commutator ensures that $k_1+k_2=k'_1+k'_2$,
\begin{eqnarray*}
  &&\hspace{-1.75em}\sin{\!\Bigl(k'_1\!\cdot\! x'_1+k'_2\cdot\! x'_2-k_1\!\cdot\! x_1-k_2\!\cdot\! x_2\!\Bigr)}\\
   &&\hspace{2em} =\sin{\!\Bigl(
           (k_1+k_2)\!\cdot\!\Bigl(\!\frac{x'_1+x'_2}{2}-\frac{x_1+x_2}{2}\!\Bigr)
                                     +(k'_1{-}k'_2)\!\cdot\! \Bigl(\!\frac{x'_1{-}x'_2}{2}\!\Bigr)
                                     -(k_1{-}k_2)\!\cdot\! \Bigl(\!\frac{x_1{-}x_2}{2}\!\Bigr)\!\Bigr)}.
\end{eqnarray*}
This factor reverses sign but is otherwise invariant under a simultaneous reflection of
$k_1+k_2\rightarrow k_1+k_2-2\frac{(k_1+k_2){\cdot}X}{X{\cdot}X}X$,
with $X=\frac{x'_1+x'_2}{2}-\frac{x_1+x_2}{2}$, and reversals of $k_1-k_2\rightarrow -(k_1-k_2)$
and of $k'_1-k'_2\rightarrow -(k'_1-k'_2)$.
$\tilde M(k_1-k_2;k'_1-k'_2)$ is required to be invariant under the simultaneous reversals
$k_1-k_2\rightarrow -(k_1-k_2)$, $k'_1-k'_2\rightarrow -(k'_1-k'_2)$, so that
$[\hat\phi(x_1,x_2),\hat\phi(x'_1,x'_2)]$ is trivial when the equally weighted center of mass separation $\frac{x'_1+x'_2}{2}-\frac{x_1+x_2}{2}$ is space--like, which is a more--or--less natural extension of Locality for the 2--particle case.

An example of a positive semi--definite infinite--dimensional matrix for
$\tilde M(k_1\!-\!k_2;k_1'\!-\!k_2')$ may be constructed as a Hadamard exponential\cite[Lemma 2.5]{Reams} $\exp(-d(k_1{-}k_2,k'_1{-}k'_2))$, in terms of an almost negative--definite distance matrix constructed using the hyperbolic metric distance\cite[\S 6.4, p.119]{DistanceEncyclopedia} between $k_1{-}k_2$ and $k'_1{-}k'_2$.
Explicitly, we could, enforcing that $k_1{-}k_2$ and $k'_1{-}k'_2$ are both on mass--shell for some mass $\Mass$ and either both timelike forward or both timelike backward, effectively enforcing invariance under reversals of $k_1-k_2\rightarrow -(k_1-k_2)$
and of $k'_1-k'_2\rightarrow -(k'_1-k'_2)$, take
\begin{eqnarray*}
  \tilde M(k_1\!-\!k_2;k_1'\!-\!k_2')&=&
\left(\rmi\tilde\Delta_{+\Mass}(k_1{-}k_2)\rmi\tilde\Delta_{+\Mass}(k'_1{-}k'_2)
    +\rmi\tilde\Delta_{+\Mass}(k_2{-}k_1)\rmi\tilde\Delta_{+\Mass}(k'_2{-}k'_1)\right)\cr
  &&          \ \times\ \exp(-d(k_1{-}k_2,k'_1{-}k'_2)),\cr
   \mbox{where}\ d(k,k')&=&\alpha\;\mathrm{arccosh}\!\left[\!\frac{k\cdot k'}{\Mass^2}\!\right]
      =\alpha\ln\left[\frac{k\cdot k'}{\Mass^2}+\sqrt{\left(\!\frac{k\cdot k'}{\Mass^2}\!\right)^{\!\!2}-1}\right],
\end{eqnarray*}
for some constant $\alpha>0$.
Expanding the exponential and logarithm and simplifying,
\begin{eqnarray*}
  \tilde M(k_1\!-\!k_2;k_1'\!-\!k_2')&=&
\left(\rmi\tilde\Delta_{+\Mass}(k_1{-}k_2)\rmi\tilde\Delta_{+\Mass}(k'_1{-}k'_2)
    +\rmi\tilde\Delta_{+\Mass}(k_2{-}k_1)\rmi\tilde\Delta_{+\Mass}(k'_2{-}k'_1)\right)\cr
  &&          \ \times\ \left[\frac{(k_1{-}k_2)\cdot (k'_1{-}k'_2)}{\Mass^2}
                     -\sqrt{\left(\!\frac{(k_1{-}k_2)\cdot (k'_1{-}k'_2)}{\Mass^2}\!\right)^{\!\!2}-1}\right]^\alpha.
\end{eqnarray*}
We can use sums of this matrix for multiple values or continuous ranges of internal mass $\Mass$, with varying values of $\alpha(\Mass)$; or, indeed, we can use any metric--preserving function\cite[\S 4.1, p.80]{DistanceEncyclopedia} applied to the hyperbolic metric distance $d(k,k')$, not just multiplication by a constant $\alpha$.

We can construct real--space equivalents of the 2--particle--to--2--particle VEVs,
\begin{eqnarray*}
  &&\VEV{\hat\phi(x_1,x_2)\hat\phi(x_1',x_2')}\cr
  &&\hspace{4em}=\!\!\int\!\!\VEV{a(k_1,k_2)a^\dagger(k_1',k_2')}
      \rme^{-\rmi\left(k_1\cdot x_1+k_2\cdot x_2-k_1'\cdot x_1'-k_2'\cdot x_2'\right)}
              \Fint{k_1^{\ }}\,\Fint{k_2^{\ }}\,\Fint{k_1'}\,\Fint{k_2'},\cr
&&\hspace{4em}= \rmi\Delta_{+\MMass}\left(\!\frac{x_1+x_2}{2}-\frac{x'_1+x'_2}{2}\!\right)
                                M\!\left(\!\frac{x_2-x_1}{2};\frac{x_1'-x'_2}{2}\!\right).
\end{eqnarray*}
The unfamiliar structure $M\!\left(\!\frac{x_2-x_1}{2};\frac{x_1'-x'_2}{2}\!\right)$ may be made to act as a kind of resonance to allow relatively large contributions to 2--particle--to--2--particle VEVs even at large time--like or space--like separation if $x_2-x_1$ is close to $x_1'-x_2'$.

The construction has so far used an \emph{equal weight} center of mass, with an internal mass $\Mass$ that is associated with the separation $x_1-x_2$, not with $x_1$ and $x_2$ separately.
For unequal weights $m_1$, $m_2$, $m_1+m_2=m$, where $\MMass$, $\Mass$, and $m$ may all be different (although only the relative weights $m_i/m$ are significant), we may write $k_1\cdot x_1+k_2\cdot x_2$ as
$$k_1\cdot x_1+k_2\cdot x_2=(k_1+k_2)\cdot\left(\!\frac{m_1 x_1+m_2 x_2}{m}\!\right)
                 +\frac{m_1m_2}{m}\left(\frac{k_1}{m_1}-\frac{k_2}{m_2}\right)\cdot(x_1-x_2)$$
and use
$$
  [a(k_1, k_2),a^\dagger(k'_1, k'_2)]=\Fdelta^4(k_1+k_2-k'_1-k'_2)
                   \rmi\tilde\Delta_{+\MMass}(k_1+k_2)
                   \tilde M\left(\frac{k_1}{m_1}-\frac{k_2}{m_2};\frac{k'_1}{m_1}-\frac{k'_2}{m_2}\right),
$$
requiring invariance of $\tilde M({k_1}/{m_1}-{k_2}/{m_2};{k'_1}/{m_1}-{k'_2}/{m_2})$ under
reversal of its two parameters, resulting in Locality being determined by whether
$X={(m_1 x_1+m_2 x_2)}/m$ is space--like separated from $X'={(m_1 x'_1+m_2 x'_2)}/m$.

Because Locality of $\hat\phi(x_1,x_2)$ is determined only by the center of mass $X=(m_1 x_1+m_2 x_2)/m$, it is appropriate to write $\hat\phi(x_1,x_2)$ in an alternative form,
$$
  \hat\phi_z(X)=\hat\phi\left(X-\frac{m_2z}{m},X+\frac{m_1z}{m}\right),\qquad\mbox{where }z=x_2-x_1.
$$
In this notation, it is clear that $\hat\phi_z(X)$ is no more than a Wightman field at $X$, with $z$ an index for a potentially elaborate nonlinear representation of the Lorentz group.

For zero weight for $x_2$, for which
$k_1\cdot x_1+k_2\cdot x_2=(k_1{+}k_2)\cdot x_1+k_2\cdot(x_2{-}x_1)$, we can use
\begin{eqnarray*}
  [a(k_1, k_2),a^\dagger(k'_1, k'_2)]&=&\Fdelta^4(k_1+k_2-k'_1-k'_2)
                   \rmi\tilde\Delta_{+\MMass}(k_1+k_2)\cr
&&\hspace{-3em}\times\ 
   \left(\rmi\tilde\Delta_{+\Mass}(-k_2)\rmi\tilde\Delta_{+\Mass}(-k'_2)
       +\rmi\tilde\Delta_{+\Mass}(k_2)\rmi\tilde\Delta_{+\Mass}(k'_2)\right)\exp(-d(k_2,k'_2)),
\end{eqnarray*}
for which whether the commutator $[\hat\phi(x_1,x_2),\hat\phi(x'_1,x'_2)]$ is trivial depends only on whether  $x'_1{-}x_1$ is space--like.
In this case, $x_2{-}x_1$ is essentially a parameter for a quantum field at $x_1$, so that an appropriate alternative form is $\hat\phi_z(x)=\hat\phi(x,x+z)$.

\subsection{The $n$--particle center of mass localized field}
\label{nParticle}
The 2--particle construction can be extended to $n$ particles,
$$
  [a(k_1,...,k_n),a^\dagger(k'_1,...,k'_n)]=\Fdelta^4(K{-}K')
                   \rmi\tilde\Delta_{+\MMass}(K)
     \tilde M\left(\!\left[\frac{k_{i+1}}{m_{i+1}}{-}\frac{k_i}{m_i}\right]_{i=1}^{\!n{-}1}\!\!\!;\!
                      \left[\frac{k'_{i+1}}{m_{i+1}}{-}\frac{k'_i}{m_i}\right]_{i=1}^{\!n{-}1\!}\right)\!\!,
$$
where $K=\sum_{i=1}^n k_i$ and $K'=\sum_{i=1}^n k'_i$.
$\rmi\tilde\Delta_{+\MMass}(K)$ is invariant under space--like reflections and we require that $\tilde M(\cdots;\cdots)$ is a positive semi-definite matrix, Lorentz invariant, and invariant under simultaneous reversal of all its parameters.
Using the identity
$$\sum_{i=1}^n \left[k_i\cdot x_i\right]=K\cdot X+\sum_{i=1}^n\sum_{j=1}^{i-1}
                                  \frac{m_i m_j}{m}\!\left(\!\frac{k_i}{m_i}-\frac{k_j}{m_j}\!\right)\cdot(x_i-x_j),$$
where $m=\sum_{i=1}^n m_i$ is the total mass and $X=\frac{1}{m}\sum_{i=1}^n m_i x_i$ is the center of mass of the $n$ particles,
we can show that $[\hat\phi(x_1,...,x_n),\hat\phi(x'_1,...,x'_n)]$ is trivial whenever the separation between the center of masses, $X'-X$, is space--like.

It is again appropriate to write $\hat\phi(x_1,...,x_n)$ in a way that emphasizes that Locality is determined purely by the center of mass $X$, using the relative separations $z_i=x_{i+1}-x_i$,
$$
  \hat\phi_{z_1,...,z_{n-1}}(X)=\hat\phi(x_1,...,x_n),
      \quad\mbox{where }
      x_n=X+\sum_{j=1}^{n-1}\left[\sum_{k=1}^j\frac{m_k}{m}z_j\right],\ 
      x_i=x_n-\sum_{j=i}^{n-1}z_j,
$$
with it again being clear that $\hat\phi_{z_1,...,z_{n-1}}(X)$ is a Wightman field at $X$, with $z_1,...,z_{n-1}$ being indices for a nonlinear representation of the Lorentz group.
Representations of the Lorentz group that can be simply expressed in a multi--particle formalism will in general not be simply expressed in terms of irreducible representations.

We can construct $\tilde M(\cdots;\cdots)$, as one of many examples, by constraining each $\frac{k_{i+1}}{m_{i+1}}{-}\frac{k_i}{m_i}$ to be time--like, using factors $\rmi\tilde\Delta_{+\Mass}\left(\pm\left(\frac{k_{i+1}}{m_{i+1}}{-}\frac{k_i}{m_i}\right)\right)$, constraining all to be forward--pointing or all to be backward--pointing, and using a Hadamard exponential of the Fujita average distance between sets\cite[Thm. 1]{Fujita} or the Hausdorff distance between sets\cite[\S 1.5, p. 47]{DistanceEncyclopedia}, using the hyperbolic metric distances $d\left(\frac{k_{i+1}}{m_{i+1}}{-}\frac{k_i}{m_i},\frac{k'_{j+1}}{m_{j+1}}{-}\frac{k'_j}{m_j}\right)$ (or using any distance matrix that is Lorentz invariant and invariant under translations of the centers of mass).

The positive semi-definite matrix $\tilde M(\cdots;\cdots)$ as a function of the $n(n{-}1)/2$ relative displacements $x_i-x_j$ and of the $n(n{-}1)/2$ relative displacements $x'_i-x'_j$ describes the response of the transition probabilities to changes of the internal structure of the measurement and preparations of an $n$--particle field, which we can think of in terms of whether the distances $d(\cdots;\cdots)$ that are used to construct $\tilde M(\cdots;\cdots)$ are large (when we obtain small transition probabilities) or small (when we obtain large transition probabilities).
For the construction as we have so far described it, we have effectively taken the distance function for different numbers of particles to be infinite, but mathematically we can extend the construction to allow the VEV $\VEV{\hat\phi(x_1,...,x_n)\hat\phi(x_1',...,x'_{n'})}$ for $n\not=n'$ to be non-zero, provided we ensure, for an empirically useful physical model, that the aggregate number of particles and anti--particles is conserved.

The $n$--particle multi--particle creation/annihilation operator commutators can be used together with the 1--particle creation/annihilation operator commutator to construct multi--particle fields using the method of \S\ref{AlmostTrivialNParticle}, however in \S\ref{MultipleCenterOfMass} we will introduce a more general construction.

\subsection{The multiple center of mass localized field}
\label{MultipleCenterOfMass}
The construction of the previous subsection can be written as
$$
  \left(\begin{array}{c}X \\ z_1 \\ \vdots \\ z_{n-1}\end{array}\right)
  =
    \left(\begin{array}{c c c c c c c}
                \mu_{1} & \mu_{2} & \mu_{3} & \cdots & \mu_{n-2} & \mu_{n-1} & \mu_{n} \\
                              -1 & 1 & 0 &\cdots & 0 & 0 & 0\\
                              0 & -1 & 1 &\cdots & 0 & 0 & 0\\
                              {} & {} & {} &\ddots & {} & {} & {}\\
                              0 & 0 & 0 &\cdots & -1 & 1 & 0\\
                              0 & 0 & 0 &\cdots & 0 & -1 & 1
          \end{array}\right)
    \left(\begin{array}{c}x_1\\ x_2\\ \vdots \\ x_{n-1} \\ x_{n}\end{array}\right),
$$
where $\sum_{j=1}^n \mu_j=1$, and for $K=\sum_{i=1}^n k_i$ there is a choice of $v_j$ for which
$$\sum_{i=1}^n \left[k_i\cdot x_i\right]=(K,v_1,\cdots,v_{n-1})\cdot
                  \left(\begin{array}{c}X \\ z_1 \\ \vdots \\ z_{n-1}\end{array}\right).
$$
We can generalize this construction by introducing an invertible linear map
$$
  \left(\begin{array}{c}X_1 \\ \vdots \\ X_m \\ z_1 \\ \vdots \\ z_{n-m}\end{array}\right)
  =
    \left(\begin{array}{c}\mu_{11} \cdots \mu_{1n} \\
                           \vdots \\ \mu_{m1} \cdots \mu_{mn} \\
                               \nu_{11} \cdots \nu_{1n} \\
                           \vdots \\ \nu_{n-m,1} \cdots \nu_{n-m,n}
          \end{array}\right)
    \left(\begin{array}{c}x_1\\ x_2\\ \vdots \\ x_{n-1} \\ x_{n}\end{array}\right),
$$
where $\mu_{ij}\ge 0$, $\sum_{j=1}^n \mu_{ij}=1$ for $i=1..m$, and $\sum_{j=1}^n \nu_{ij}=0$ for $i=1..n-m$, so that the $X_i$ are contained in the convex hull generated by the $x_i$ and transform like $x_i$ under translations and the $z_i$ are invariant under translations.
For some choice of $K_j$ and $v_j$ we require $\sum_{j=1}^m K_j=\sum_{i=1}^n k_i$ and
$$\sum_{i=1}^n \left[k_i\cdot x_i\right]=(K_1,\cdots K_m,v_1,\cdots,v_{n-m})\cdot
         \left(\begin{array}{c}X_1 \\ \vdots \\ X_m \\ z_1 \\ \vdots \\ z_{n-m}\end{array}\right),
$$
so that we can construct the $n$--particle commutator, using the method of \S\ref{AlmostTrivialNParticle}, as
$$
  [a(k_1,...,k_n),a^\dagger(k'_1,...,k'_n)]=\prod_{j=1}^m\left[\Fdelta^4(K_j{-}K'_j)
                   \rmi\tilde\Delta_{+\MMass_i}(K_j)\right]
     \tilde M\left(\!\Bigl[v_i\Bigr]_{i=1}^{\!n{-}m}\!;\!
                      \Bigl[v_i'\Bigr]_{i=1}^{\!n{-}m\!}\right)\!,
$$
where again we require that $\tilde M(\cdots;\cdots)$ is a positive semi-definite matrix, Lorentz invariant, and invariant under simultaneous reversal of all its parameters, and we can show that  $[\hat\phi(x_1,...,x_n),\hat\phi(x'_1,...,x'_n)]$ is trivial whenever the separations between the multiple centers of masses, $X'_i-X_i$, are all space--like, which will certainly be the case if the convex hulls of $x_1,...,x_n$ and of $x_1',...,x_n'$ are space--like separated.

\subsection{Invariance under internal symmetry}
\label{InvarianceUnderSymmetry}
\newcommand\Vbar{{\hspace{0.1ex}\rule[-0.3ex]{0.3ex}{2.3ex}\hspace{0.2ex}}}
For all multi--particle constructions, we can introduce a compact symmetry group $G$ that acts on parameters within the construction, so that we can introduce a multi--particle field that is invariant under the symmetry group by integration under the Haar measure,
$$\hat\phi_G(x_1, ..., x_n)=\int_G\!\hat\phi(x_1, ..., x_n\Vbar g)\mathrm{d}\mu(g),$$
where we can require for the generic multi--particle case that
$$[a(k_1, ..., k_n\Vbar g),a^\dagger(k'_1, ..., k'_n\Vbar g')]
    =\delta(g-g')C(k_1, ..., k_n; k'_1, ..., k'_n\Vbar g).
$$
For the 1--particle Klein-Gordon propagator, the only parameter is mass, whereas more elaborate group actions are possible for $n$--particle propagators, depending on their internal structures.
Smoothing the 1--particle propagator with a continuous mass distribution results in generalized free fields, which satisfy the Wightman axioms but do not have a particle interpretation\cite[\S 3.4]{Streater}, however smoothing over internal masses and over other parameters may be useful for $n$--particle fields.

\section{Dirac and Maxwell multi--particle fields}
\label{DiracMaxwellFields}
\newcommand{\kD}{{k}}\newcommand{\xD}{{x}}
\newcommand{\kM}{{u}}\newcommand{\xM}{{y}}
The introduction of Lorentz covariant internal degrees of freedom introduces significantly more structural possibilities, including the requirement that Dirac 1--particle fields (and, in general, Dirac $n$-particle fields for odd $n$) must anticommute to satisfy Locality.

The interaction--free Dirac and Maxwell 1--particle fields can be presented as
\begin{eqnarray*}
  \hat\psi_a(x)&=&\int\!\left[b_a(k)\rme^{-\rmi k\cdot x}
                                   +d^c_a(k)\rme^{\rmi k\cdot x}\right]\Fint{k},
      \qquad b_a(k)|0\rangle=\overline{d^{{}^{\scriptstyle c}}_a(k)}|0\rangle=0,\cr
  \{b_a(k),\overline{b_{a'}(k')}\}
           &=&\Fdelta^4(k-k')(\gamma\cdot k+\Mass)_{aa'}\rmi\tilde\Delta_{+\Mass}(k),\cr
  \{d^c_a(k),\overline{d^c_{a'}(k')}\}
           &=&\Fdelta^4(k-k')(\gamma\cdot k-\Mass)_{aa'}\rmi\tilde\Delta_{+\Mass}(k),\cr
\{\hat\psi_a(x),\overline{\hat\psi_{a'}(x')}\}&=&\!\int\!\!\left[
  \{b_a(k),\overline{b_{a'}(k')}\}\rme^{-\rmi k\cdot x+\rmi k'\cdot x'}
  \!\!+\!\{d^c_a(k),\overline{d^c_{a'}(k')}\}\rme^{\rmi k\cdot x-\rmi k'\cdot x'}\right]\!\Fint{k}\,\Fint{k'}\cr
  &=&\int(\gamma\cdot k+\Mass)_{aa'}\left[\rmi\tilde\Delta_{+\Mass}(k)-\rmi\tilde\Delta_{+\Mass}(-k)\right]
                         \rme^{-\rmi k\cdot (x-x')}\Fint{k},\cr
  &=&(\rmi\gamma\cdot\partial_x+\Mass)_{aa'}\rmi\Delta_m(x-x').\cr
  \rule{0pt}{5ex}\hat A_\mu(x)&=&\int\!\left[a_\mu(k)\rme^{-\rmi k\cdot x}
                                   +a^\dagger_\mu(k)\rme^{\rmi k\cdot x}\right]\Fint{k},
      \qquad a_\mu(k)|0\rangle=0,\cr
  [a_\mu(k),a^\dagger_{\mu'}(k')]
         &=& -\,\Fdelta^4(k-k')g_{\mu\mu'}\rmi\tilde\Delta_{+0}(k),\cr
  [\hat A_\mu(x),\hat A_{\mu'}(x')]&=&-g_{\mu\mu'}\rmi\Delta_0(x-x').
\end{eqnarray*}
For the Dirac field as presented here, avoiding the introduction of a basis of plane wave solutions for the Dirac equation, $d^c_a(k)$ is a creation operator and, correspondingly, $\overline{d^{{}^{\scriptstyle c}}_a(k)}$ is an annihilation operator.
The Dirac field anticommutator above reproduces \cite[Eq.\,3\raisebox{0.6ex}{\rule{1ex}{0.22ex}}170]{IZ}.
To ensure that the Maxwell field commutator is a positive semi--definite inner product, we might restrict test functions to be the divergence of a bivector, $\frac{\partial}{\partial x^\alpha}\mathfrak{f}^{[\alpha\mu]}(x)$, which, being equivalent to using the 2--form $\hat F_{\alpha\mu}=\frac{\partial}{\partial x^{[\alpha}}\hat A_{\mu]}(x)$ with the test function $\mathfrak{f}^{[\alpha\mu]}(x)$ (assuming appropriate boundary conditions at infinity), also ensures $U(1)$--gauge invariance.
If we take this approach, we can write
$[\hat F_\mathfrak{f},\hat F_\mathfrak{g}]
   =[\hat A_{\delta\mathfrak{f}},\hat A_{\delta\mathfrak{g}}]=(\delta\mathfrak{f},\delta\mathfrak{g})$.

As examples of a Dirac--Maxwell multi--particle field, we can introduce a 1--Dirac+1--Maxwell--particle field,
$$
  \hat\psi_{a\mu}(\xD,\xM)=
    \int\!\left[\mathsf{b}_{a\mu}(\kD,\kM)\rme^{-\rmi\kD\cdot\xD-\rmi\kM\cdot\xM}
    +\mathsf{d}^c_{a\mu}(\kD,\kM)\rme^{\rmi\kD\cdot\xD+\rmi\kM\cdot\xM}\right]
            \Fint{\kD}\,\Fint{\kM},
$$
either almost trivially, with
\begin{eqnarray*}
  \{\mathsf{b}_{a\mu}(\kD,\kM),\overline{\mathsf{b}_{a'\mu'}(\kD',\kM')}\}
     &=&\{\mathsf{b}_{a}(\kD),\overline{\mathsf{b}_{a'}(\kD')}\}[a_\mu(u),a^\dagger_{\mu'}(u')],\cr
  \{\mathsf{d}^c_{a\mu}(\kD,\kM),\overline{\mathsf{d}^c_{a'\mu'}(\kD',\kM')}\}
     &=&\{\mathsf{d}^c_{a}(\kD),\overline{\mathsf{d}^c_{a'}(\kD')}\}[a_\mu(u),a^\dagger_{\mu'}(u')],
\end{eqnarray*}
or relatively nontrivially, using, for example, the 2--particle propagator of an equal weight center of mass localized scalar field,
\begin{eqnarray*}
  \{\mathsf{b}_{a\mu}(\kD,\kM),\overline{\mathsf{b}_{a'\mu'}(\kD',\kM')}\}
     &=&\Bigl(\gamma{\cdot}(k{+}u)+\MMass\Bigr)_{aa'}\Bigl(-g_{\mu\mu'}\Bigr)\cr
      &&\hspace{3.5em}\times\quad
                      \Fdelta^4(k{+}u{-}k'{-}u')\rmi\tilde\Delta_{+\MMass}(k{+}u)\tilde M(k-u;k'-u')\cr
  \{\mathsf{d}^c_{a\mu}(\kD,\kM),\overline{\mathsf{d}^c_{a'\mu'}(\kD',\kM')}\}
     &=&\Bigl(\gamma{\cdot}(k{+}u)-\MMass\Bigr)_{aa'}\Bigl(-g_{\mu\mu'}\Bigr)\cr
      &&\hspace{3.5em}\times\quad
                      \Fdelta^4(k{+}u{-}k'{-}u')\rmi\tilde\Delta_{+\MMass}(k{+}u)\tilde M(k-u;k'-u'),
\end{eqnarray*}
or we can use whatever weights or other constructions are useful for empirical models.
There is no immediately obvious construction that implements the $U(1)$ and other gauge symmetries that have been so effective when using Lagrangian and other dynamical methods.
There may be a way to implement gauge symmetries in $n$--particle quantum fields formalisms, however empirical efficacy may in any case be possible without there being very close connections with contemporary methods.

\section{Generation of multilinear fields by polarization}
\label{Polarization}
Insofar as we can freely choose test functions, we can use such constructions as
$\hat\phi_{f\otimes f}$, introducing thereby a form of nonlinear dependence on test functions.
Indeed we can immediately consider constructions of homogeneous degree such as $$\hat\xi_f=\hat\phi_{f^2}+\hat\phi_{f\otimes f}+\hat\phi_f\hat\phi_f$$
and we can apply derivations Lorentz invariantly to such terms, so we can also add terms such as
$$\hat\phi_{{\partial_\mu}f{\partial^{\mu\!}}f}
    +\hat\phi_{{\partial_\mu}f\otimes {\partial^{\mu\!}}f}
    +\hat\phi_{{\partial_\mu}f}\hat\phi_{{\partial^{\mu\!}}f}$$
(we have here taken constant multipliers to be 1 wherever they could be introduced).
Given a homogeneous construction $\hat\xi_f$ of degree $n$, moreover, we can construct a multilinear quantum field operator that is symmetric in its arguments by polarization, so that homogeneous nonlinear constructions may be used to generate symmetric multilinear constructions.
For $\hat\xi_f$ as given above, we can construct
$$\hat\xi_{(2)f_1\otimes f_2}=\left.\frac{1}{2!}\frac{\partial}{\partial\lambda_1}\frac{\partial}{\partial\lambda_2}\hat\xi_{\lambda_1 f_1+\lambda_2 f_2}\right|_{\lambda_1=\lambda_2=0}.$$

We can generalize this generated construction of multilinear quantum field operators in a way that can be engineered to match rather closely with the usual construction of interacting quantum fields by
introducing a deformation of a 1--particle quantum field (here without time--ordering, even though time--ordering can be defined for both $n$-particle constructions of \S\ref{RealScalarFields}, either propagator by propagator or with time--ordering depending only on the center--of--mass),
$$\hat\xi_f=\left[\rme^{-\rmi\hat{\mathcal L}_f}\right]^\dagger
          \hat\phi_f\rme^{-\rmi\hat{\mathcal L}_f}.$$
The deformation operator $\hat{\mathcal L}_f$ is constructed using only parameterized functionals $f_\kappa$ of the test function $f$ for which $\mathrm{Supp}[f_\kappa]\subseteq\mathrm{Supp}[f]$, so that $\hat\xi_f$ is localized in the support of $f$.
We can construct as a relatively elementary example,
$$\hat{\mathcal L}_f=\int\left(\hat\phi_{f_\kappa\otimes f_\kappa}-\hat\phi_{f_\kappa}^2\right)^2
                   \rho(\kappa)\Intd\kappa,\qquad f_\kappa(y)=f(y)+ \kappa f(y)^3.$$
This construction effectively ensures that the test function $f$ generates 2--particle operators $\hat\phi_{f_\kappa\otimes f_\kappa}$ for all $\kappa$ in the support of $\rho(\kappa)$, depending at the lowest level of expansion on the intensity of the crossover given by the bilinear form $(f^*_\kappa, f)$ and at higher levels of expansion also on $(f^*_{\kappa'}, f_\kappa)$ for different values of $\kappa$, $\kappa'$.

Given such constructions of $\hat\xi_f$, which are nontrivially dependent on $f$, we can generate a system of $n$--particle multilinear quantum field operators by polarization for any $n$,
$$\hat\xi_{(n)f_1\otimes\cdots\otimes f_n}=\left.\frac{1}{n!}
            \left[\prod_{i=1}^n\frac{\partial}{\partial\lambda_i}\right]\hat\xi_{\sum\lambda_i f_i}\right|_{\underline{\lambda}=0}.$$
At the lowest level, we always have $\hat\xi_{(1)f}=\hat\phi_f$.
$\big[$If $\hat\xi_{\sum\lambda_i f_i}$ is dependent on the $\lambda_i^*$ as well as on the $\lambda_i$, as may well be the case, then we can also generate multiparticle quantum field operators using derivatives $\partial/\partial\lambda_i^*$.$\big]$
We may be able to reconstruct $\hat\xi_f$ as
$$\hat\xi_f=\sum\limits_{n=1}^\infty\hat\xi_{(n)f^{\otimes n}},$$
if this expression exists, which seems, however, quite unlikely given our experience with interacting quantum fields.
If a Hilbert space \emph{can} be constructed that supports a cyclic representation of $\hat\xi_f$, then we could construct a cyclic representation of $\hat\xi_{(n)f_1\otimes\cdots\otimes f_n}$ by polarization, so we could equivalently take the Hilbert space to support a cyclic representation of $\hat\xi_{(n)f^{\otimes n}}$.
Otherwise, however, we can nonetheless take $\hat\xi_f$ to be a \emph{formal} generating function for the multilinear operators $\hat\xi_{(n)f_1\otimes\cdots\otimes f_n}$, then, for example, we can use the vacuum state over the algebra generated by $\hat\xi_{(n)f^{\otimes n}}$, all of which are functions of whatever Gaussian fields are used in the construction of $\hat{\mathcal L}_f$, to construct a vacuum Fock--Hilbert space.

$f_\kappa$ can be any functional of $f$, and $\kappa$ can be taken from any parameter space.
In particular, we can take $\kappa$ to be a position in space--time, in which case we can
construct an operator that is rather close to a conventional interaction term, using a method that might be called ``test function regularization'', such as
$$\hat{\mathcal L}_f=\int\hat\phi_{f_x}^4\Intd^4x,\qquad f_x(y)=|f(y)\cdot f(x+\mu(f)(y-x))|^2,$$
and introducing time--ordering,
$$\hat\xi_f=\mathrm{T}\left[\rme^{-\rmi\hat{\mathcal L}_f}\right]^\dagger
          \mathrm{T}\left[\hat\phi_f\rme^{-\rmi\hat{\mathcal L}_f}\right].$$
$f_x(y)$ can be constructed in very many different ways, under the conditions that ---as for the example given, provided the real scalar value $\mu(f)$ increases sufficiently fast--- at a given point $x$, $\hat\phi_{f_x}$ is localized in the support of $f$ because $\mathrm{Supp}[f_x]\subseteq\mathrm{Supp}[f]$; and $\hat\phi_{f_x}$ is also localized increasingly close to the point $x$ as $f$ becomes closer to the plane waves used when we construct the S--matrix of the theory, so that $\hat\phi_{f_x}$ becomes ``close'' to $\hat\phi(x)$ and $\hat{\mathcal L}_f$ becomes ``close'' to the conventional $\hat\phi^4$ Lagrangian interaction.
Although $f_x(y)$ in general does not approach $\delta^4(x-y)$ in a Lorentz invariant way, renormalizability of a theory is perhaps exactly the property of being insensitive to different regularizations when parameters of the Lagrangian are taken to flow appropriately, in this case as a function of $\mu(f)$; for nonrenormalizable theories, for which regularization does make a difference, the nonlinear dependence of $f_x$ on $f$ fixes a regularization precisely, albeit rather intractably.
The construction by polarization of the multilinear field $\hat\xi_{(n)f_1\otimes\cdots\otimes f_n}$, using the nonlinear dependence of the test function regularized $\hat\xi_f$ on $f$, might in traditional terms be called an expansion in powers of the test function $f$.

Taking the test function $f$ to be a plane wave, which is fundamental to the idea of an S--matrix approach to phenomenology, is typically used as an idealized description of an experiment, however experiments in fact do not prepare states that are extended over all space, so a detailed description should include how the test function is localized.
We note also that constructions of this historically motivated type may not provide the most succinct and tractable physical models, given that so many structurally very different alternatives are available that use general parameter spaces and that use 2--particle or higher Gaussian fields in the construction of $\hat{\mathcal L}_f$.

\section{Discussion}
Part of the motivation here is that it is worthwhile to present the systematics of phenomena as an alternative to and without necessarily knowing or understanding the (quantum) dynamics that cause the systematics, even though we might prefer to know and understand the dynamics.
A wide range of multi--particle models is possible ---much extending the modeling reach of quantum optics, for example--- however such models will be more useful if the range of models is constrained enough that almost all models can be physically realized.
A theory that includes models that cannot be physically realized may still be useful, but it will be significantly more useful if there is a more--or--less clear path to realization for almost all of the theory's models; it will be important to restrict the allowed structure of multi--particle models as far as possible by the application of symmetries or other constraints.

A clear distinction can be made between Gaussian and non--Gaussian $n$--particle quantum fields.
It is not necessary to introduce large $n$ Gaussian quantum fields to enable the construction of large $n$ non--Gaussian quantum fields, however if large $n$ Gaussian quantum fields are introduced
then it is not necessary that the $n$--particle two operator VEV has to be even remotely similar in construction to the $m$--particle two operator VEV for $m\not=n$, corresponding to the emergence of entirely novel properties as the number of particles increases.
If large $n$ Gaussian quantum fields \emph{are} introduced, we can hope for the constraint of a relatively straightforward, empirically effective formula that generates $n$--particle two operator VEVs for all $n$.
In contrast, we have seen that the generation of non--Gaussian $n$--particle quantum fields can follow the familiar construction of interacting quantum fields quite closely.

The usual presentation of an interacting quantum field is in terms of a quantized classical field dynamics that is expected to cause the systematics of multi--particle dynamics, but in that case we also have to present a renormalization scheme, which in general is a more elaborate object than the interaction terms.
The multi--particle constructions suggested here are perhaps no more extraordinary as an extension of the 1--particle dynamics, with the advantage that they are mathematically better--defined.

A test function perspective can be taken to motivate a stochastic signal processing interpretation of quantum field theory, in which powers of $\hat\phi_g$ (and sums of powers) can be understood to \emph{stochastically modulate} the vacuum state vector,
$|0\rangle\mapsto\hat\phi_g|0\rangle$,
$|0\rangle\mapsto\hat\phi_g^2|0\rangle$, ..., or other vectors,
$\hat A|0\rangle\mapsto\hat\phi_g\hat A|0\rangle$,
$\hat A|0\rangle\mapsto\hat\phi_g^2\hat A|0\rangle$, ..., with the vacuum state being a stochastically nontrivial, Poincar\'e invariant starting point.
Any effective contemporary physical theory presumably must be a stochastic signal processing formalism, insofar as the complete set of the electronic signals generated by all our experiments and stored in a lossily compressed digitized form in computer memory (ultimately including digitized photographs, videos, and all), or at least the set of statistics we generate from some chosen part of that data, is the explanandum of the theory.
In a classical signal processing context, it would hardly be expected that a linear response to a given modulation would be better than a first approximation; a multi--particle, multi--point, multilinear formalism is one way to provide additional resources to describe and predict nonlinear responses to a given modulation.
One consequence for experiment of a stochastic signal processing interpretation is that how much a measured response depends nonlinearly or multilinearly on many details of a given modulation, not just on the isolated detail of an average wave--number, becomes a question of greater than usual experimental interest that suggests a focus on high precision as much as or more than on high energy.

The approach here is essentially elementary, using a notation and ideas that are very little removed from those of the Wightman axioms and formalizing the usual emergence of quasiparticles and collective excitations.
We have done no more than to allow a quantum field to be an operator--valued distribution $\hat\phi(x_1, ..., x_n)$ instead of $\hat\phi(x)$ and to accommodate this change as naturally as seems possible in the other axioms.
In terms of test functions and systems of nonlinear stimuli and responses, we have suggested we might more use the unexceptional constructions $\hat\phi_f^n$ and $\hat\phi_{f^n}$, and we have introduced $\hat\phi_{f^{\otimes n}}$ in two forms, as the almost trivial multi--particle fields of \S\ref{AlmostTrivialNParticle} and as center of mass localized fields, all of which may be used or not in the construction of physically useful models or approximations to other models.
The $\hat\phi_{f^{\otimes n}}$ satisfy forms of locality that may seem too weak, but gradual changes to our idea of locality have been a consistent aspect of quantum mechanics over the years.
There are perhaps better formalisms for augmenting the free Wightman field Fock--Hilbert space, nonetheless the constructions given above of $n$--particle quantum fields suggest a new direction and a new inspiration for model building regardless of formalism.


\begin{thebibliography}{99}
\bibitem{Fraser} Fraser, D. (2006). Haag's Theorem and the Interpretation of Quantum Field Theories with Interactions. Ph.D. thesis. U. of Pittsburgh. [http://d-scholarship.pitt.edu/8260/].

\bibitem{EarmanFraser} Earman, J. \& Fraser, D. (2006). Haag's Theorem and its Implications for the Foundations of Quantum Field Theory. Erkenntnis 64, 305.

\bibitem{Haag}Haag, R. (1996). Local Quantum Physics, 2nd Edn. Springer, Berlin.

\bibitem{FredenhagenEtAl}Fredenhagen, K., Rehren, K.-H., \& Seiler, E. (2007). Lect. Notes Phys. 721, 61–87. Springer, Berlin.

\bibitem{Streater}Streater, R. F. (1975). Rep. Prog. Phys. 38,771.

\bibitem{Summers}Summers, S. J. (2012). arXiv:1203.3991v1 [math-ph].

\bibitem{Reams}Reams, R. (1999). Linear Algebra and its Applications, 288, 35.

\bibitem{DistanceEncyclopedia}Deza, M. M. \& Deza, E. (2009). Encyclopedia of Distances. Springer, Berlin.

\bibitem{Fujita}Fujita, O. (2013) Japan J. Indust. Appl. Math., 30, 1.

\bibitem{IZ}Itzykson, C. \& Zuber, J.--B. (1980). Quantum Field Theory. McGraw--Hill, New York.



\end{thebibliography}
\end{document}